# Dynamical Effects on Driven Dislocation Glide through Local Pinnings

Masato Hiratani and Vasily V. Bulatov
Chemistry and Materials Science Directorate, Lawrence Livermore National Laboratory, Livermore, CA 94550

We present effects of dislocation inertia on the driven dislocation glide through local immobile pinnings using a stochastic computational model. The global dislocation velocity at a higher stress range is found noticeably dependent on the dislocation inertia, and the temperature sensitivity is observed to be strongly non-Arrhenius. The statistical analysis indicates that the correlation of the local dislocation kinetic energy is extended at a lower temperature, which results in the enhanced depinning rate by the inertia effect.

Percolation of quantized plastic carriers, dislocations, through local disorders has been intensively investigated for long time due to its significant importance in engineering application of strengthening materials. It has also attracted the attention for its manifestation of various complex dynamical phases [1], commonly pertaining to system of line defects with topological charges, such as fluxoids in superconductors [2], or vortices in superfluids [3]. Many of these studies are usually focused on the overdamped cases, and the kinetic energy of the defects has been largely neglected.

Consider a simple model for an early deformation stage of well annealed dilute alloys. There each dislocation is well separated from others but it encounters random arrays of immobile local pinnings during its glide. Under a constant applied force $F$, individual dislocation moves certain distance, and eventually it gets trapped by the surrounding obstacles with a certain local pinning force $F_p$ due to their collective pinning effect when the driving force does not exceeds the percolation threshold $F_c$. At finite temperature $T > 0$, dislocations can maintain the forward motion macroscopically as the sequence of the thermally induced depinning and the flight to the next metastable configurations. At a low temperature and a low drive, the time duration $t_{th}$ for the thermal activation at metastable pinning configurations is much longer than the flight time $t_{fl}$ between them, and dislocation motion is often described as hops between them. In such a case, the temperature dependence of the dislocation velocity $v$ follows the Arrhenius law as $v \propto \exp(-G/kT)$, and the activation enthalpy $G$ has a strongly non-linear drive dependence as $G \propto (F_c/F)^\mu$ that diverges in the limit of small drive [4]. As the drive increases toward the critical, the activation enthalpy is reduced by the external mechanical work, and shorter $t_{th}$ can be comparable to $t_{fl}$ particularly when the kinetic damping $B$ is high. In this case, the flight process is not negligible anymore, and the $t_{fl}$ has to be included into the elapsed travel time. Provided that these processes are independent, one could estimate the average dislocation velocity as $v = \langle \Lambda/(t_{th}+t_{fl}) \rangle$ where $\Lambda$ is the interval between the successive metastable configurations. While the flight velocity in the steady motion is determined by the $B$ and the local driving force $f$, the effective dislocation mass $m$ determines the transient dislocation dynamics during the time scale of the relaxation $\tau_o \sim 2m/B$. For instance, in the framework of Granato-Lücke dislocation string model [5], a single dislocation dynamics is described with an attenuated wave equation, for which general solution is obtained by using a kernel [6] as

$$\psi(x,t) = (1/mc)\int_{-\infty}^{t} dt' \int_{x-ct}^{x+ct} dx' \, f(x',t') K(x'-x, t'-t) \quad (1)$$

where $K(x,t) \equiv I_o\left(\sqrt{(t/\tau_o)^2 - (x/c\tau_o)^2}\right) e^{(-t/\tau_o)}$, $c$ the speed of the shear wave, $I_o$ the modified Bessel function of the first kind of the order zero. Due to the finite extent of the memory described by the kernel, the correlated dislocation response is seen even when it's subject to the uncorrelated random local force, e.g. in case of white noise, we obtain

$$\langle \psi(x,t)\psi(x+dx, t+dt) \rangle / \langle \psi(x,t)^2 \rangle = \int_{u/2}^{\infty} d\tau \int_{-c\tau}^{c(\tau-u)} d\xi \, K(\xi,\tau) K(\xi+dx, \tau+dt), \quad (2)$$

where $u = (dx/c - dt)\cdot \theta(dx/c - dt)$ and $\theta(x)$ is the step function. Similarly, the finite correlation can be derived for the local velocity or the kinetic energy at two different positions and times as well. The time scale of the $\tau_o$ has been recognized to be small as compared with the $t_{th}$ or $t_{fl}$, but it may be long enough for the high velocity spots on the dislocation line to overshoot the metastable position significantly, or to propagate over the length scale of local pinnings. Thereby, the activation enthalpy may be altered during the time scale by the transient dynamics of the flight process [7].

Some experimental studies evidence the alternation of the macroscopic plasticity due to the dislocation inertia. Cryogenic experiments demonstrated the significant reduction of the dislocation velocity by switching on the magnetic field while the applied load and temperature were kept to be same: the electron drag was effectively enhanced due to the dissociation of Cooper pairs in superconductors [8] or the induction of cyclotron motion in [9]. Since the velocity change is so large, typically ranging over 2~3 orders of magnitude, that it cannot be described solely by the change in the drag and the flight time [9].

In fact, athermal computational experiments also exhibited the inertia effect as the hysteresis of velocity-flow stress relation [10, 11], but the models are not applicable to the situation where both the flight process and the thermal activation process are inneligible.

In the face-centered cubic structured metals such as copper, one of the major contribution to the damping $B$ at a not very low temperature is phonon-drag [12]. The phonon drag is expected to decrease as temperature decreases; therefore, there is a competition between the thermal activation process and flight process in this system due to the opposite temperature sensitivity. In addition, the relaxation time $\tau_o$ becomes longer at a lower temperature and the transient dislocation behavior can be significant when dislocations interact with pinning obstacles, e.g. local hot spots on the dislocation propagate over the activation length. This spatio-temporal non-locality of the dynamical effect can be measured with the correlation of dislocation kinetic energy fluctuation:

$$C(d\mathbf{x},dt) = \langle [E(\mathbf{x},t) - \langle E(\mathbf{x},t) \rangle][E(\mathbf{x}+d\mathbf{x},t+dt) - \langle E(\mathbf{x}+d\mathbf{x},t+dt) \rangle] \rangle . \quad (3)$$

Obviously, the aforementioned stress-temperature regime cannot be accessible by previously studied deterministic models, nor overdamped stochastic models, and has been rarely explored so far. In this Letter, thermally agitated underdamped dislocation motion is investigated computationally using the stochastic dislocation dynamics scheme. We report a direct evidence of the dislocation inertia effect on the global dislocation dynamics.

In the scheme of discrete dislocation dynamics (DD), all dislocation lines are typically represented as linkages of shorter dislocation line elements, and pairwise interactive forces are computed according to the line element directions and Burgers vectors [13-15]. Here we consider the Langevin DD model for a pair of Shockley partial dislocations. The stochastic equation of motion for the i-th dislocation element is taken as

$$\mathbf{M}_i \cdot \frac{d^2 \mathbf{r}_i}{dt^2} = -\mathbf{B}_i \cdot \frac{d \mathbf{r}_i}{dt} + \sum_q \mathbf{f}_i^q + \mathbf{\Gamma}_i, \quad (4)$$

where $\mathbf{M}_i$ is the effective mass density tensor, $\mathbf{B}_i$ the total drag tensor, $\mathbf{f}_i^q$ the various deterministic force components per unit length, which consist of the dislocation-dislocation interactive force, the short range obstacle pinning force, the surface force due to the stacking fault, the image force according to the boundary condition, the external force. Here identical dilatation centers are used as the local pinnings. In addition, anisotropy of the mass and the drag is neglected here, and the dislocation self-energy $E_i^{self}$ are approximated as $E_i^{self} \approx \mu b_i^2 / 2$ to estimate the effective mass density $m_i = E_i^{self} / c^2$ [16] where $b_i$ is the Burgers vector of the partial dislocation element and $\mu$ is the shear modulus of the elastic media. All the $\mathbf{f}_i^q$ is computed based on the isotropic linear elasticity [17]. The last term in Eqn (3) denotes the randomly sampled thermal agitation modeled as a white noise with a zero mean and a correlation of $\langle \mathbf{\Gamma}_i(0) \cdot \mathbf{\Gamma}_j(t) \rangle = 2kTB(T)\delta_{ij}\delta(t)$ [18]. Distance is scaled by the magnitude of the perfect Burgers vector $b$, and dislocation lines are discretized into elements of the average length of $\Delta l = 5b$. The element properties are updated every $\Delta t = 100$ fs. Material parameters are chosen for copper using values available in atomistic studies and experiments [19]. All simulations are performed in a cubic cell of size of $1600b$ bounded by $(1\bar{2}1), (\bar{1}01)$ and $(111)$ faces. A pair of Shockley partial dislocations stemmed from an edge dislocation $1/2[\bar{1}01] = 1/6[\bar{1}\bar{1}2] + 1/6[\bar{2}11]$ is introduced along $\langle 1\bar{2}1 \rangle$ on a (111) plane under the applied shear stress of $\sigma_{[111][\bar{1}01]}$. Periodic boundary conditions are imposed on the all faces, and the image force is evaluated to correct the force field around each dislocation element in the primary cell [20].

At simulated temperatures between 200K and 600K, thermalization towards the assigned temperature of heat bath is achieved within 20 ps. Figure 1a shows the swift transition of the dislocation kinetic temperature $T_k$ after the Langevin force is turned on at $t = 0$. The partial dislocations are placed on the glide plane free from the external stress and local disorders at the heat bath temperature of 400K. We also observe the convergence of the velocity profile toward the Maxwell-Boltzmann distribution during the same time scale. In the same relaxed system, within the all simulated temperatures, the dissociation width is seen to oscillate around an equilibrium distance $d_o \sim 36$ Å (Fig. 1b). A similar oscillation of the stacking fault width known as dislocation breathing is reported for a screw dislocation in a study by molecular dynamics [21]. The larger dissociation width and the longer breathing period are obtained for the edge dislocation compared with a screw in the DD simulations. The period of major peaks in the simulation data of the edge is obtained to be $t_o \sim 11$ ps, close to a theoretical value of $\pi\sqrt{2md_o/\gamma} \sim 12.4$ ps [21] where $\gamma$ is the stacking fault energy.

The global dislocation behavior in the presence of local disorder, over 2 ns is shown in the Fig. 2 ('w mass' data). The data is obtained at 200 K and 600 K below the threshold stress, and averaged over 2ns. Identical dilation centers are randomly distributed on a (111) plane above a parallel glide plane with the average inplane spacing of $L = 25b$. By using athermal force balance scheme, the pinning force of individual obstacle against a pair of the partial dislocations is measured to be $F_p$=210±10 pN, and the critical resolved shear stress required to percolate through the random obstacle arrays is determined to be 110±5 MPa. At finite temperature simulations, the dislocations move forward in jerky manner below the critical stress, and the average velocity follows a non-linear stress dependence. At lower stress, the dynamics is mainly controlled by duration of the thermal activated depinning, and thus, temperature sensitivity of the velocity is positive, as typical of the Arrehenius type dependence. As stress increases, the role of the flight process becomes larger and noticeable, seen as mitigation of the stress dependence in the velocity data. Due to the negative temperature sensitivity of the flight time, temperature sensitivity of the velocity is reduced, becomes zero, and reversed as stress increases. Such a crossover of the two processes is observed at $\sigma_x$ ~75 MPa and $v_x$ ~ 49 m/s.

To make evident the inertia effect, simulations with the overdamped stochastic scheme are also exercised by dropping a left hand term of Eqn. 4 (Fig.2, 'wo mass' data). Noticeble velocity reduction is indeed seen at higher stress at both temperatures by virtually suppressing the inertia, for example at the stress of 95 MPa, the velocity changes from 141 m/s to 95 m/s at 200K, and from 97 m/s to 79 m/s at 600 K. The onset stress at which the velocity bifurcation takes off seems to be higher at 600 K. In general, the smaller velocity change is noted at lower stress and at the higher temperature.

The difference in the global velocity behavior is consequence of the change in the dynamical interaction between the dislocations and local pinnings. To better understand this change, the velocity data at the crossover point $\sigma_x$ is analyzed. Figure 3 displays the local velocity profile at the $\sigma_x$. Although the net contribution from the thermal activated depinning process and the flight process is same at the two different temperatures, the internal velocity profile is considerably different. At 600 K, the propagation of the high velocity spot is disrupted and limited within the certain time and extent along the dislocation, at variance with the diagonal streak pattern at 200 K suggests the longer propagation of the hot spot. Longer propagation at 200 K is related to the longer relaxation time i.e. $\tau_o$ = 20.1 ps for the perfect dislocation while most of the hot spot dies out before it reaches adjacent pinning site at 600K for the shorter relaxation time of 11.4 ps.

Similar indication can be found in the dynamical correlation of the kinetic energy fluctuation $C(dx, dt)$ (Fig. 4). Short correlation at 600 K as compared with the average obstacle spacing of $L = 25b$ implies that the kinetic energy transfer to the distant pinning sites is less likely and the inertia effect on the depinning process is limited. At 200 K, on the other hand, the correlation is extended. Hence, the depinning process is non-local due to the coupling with the transient flight process. Furthermore, this can make individual thermal activation events more correlated and non-Markovian.

Current study is limited to the single dislocation glide. Further investigation is needed to evaluate the inertia effects for many dislocation system where the quick relaxation cannot be expected due to the long range interaction.

In conclusion, we have examined the thermally and mechanically driven dislocation motion through local pinnings by employing the stochastic discrete dislocation dynamics scheme. The noticeable contribution of the dislocation inertia to the average dislocation velocity is confirmed by the simulations. The underdamped dislocation flight process becomes progressively more essential as increasing stress toward the threshold stress lowering the unpinning activation enthalpy through the kinetic energy transfer.

Authors greatly appreciate Dr. R. Rudd for valuable suggestions. This work was performed under the auspices of the U.S. Department of Energy by the University of California, Lawrence Livermore National Laboratory under Contract No. W-7405-Eng-48, and the support of the DOE (Grant No. DE-FG03-01ER54629).

Figure Captions/Labels

Fig.1 a (top) fluctuation of Shockley interval, b (bottom) dislocation kinetic temperature as function of time.

Fig.2 Average dislocation velocity as function of the resolved shear stress at 200 K and 600 K. Simulation data is obtained by taking the dislocation mass (w mass), and without mass (wo mass).

Fig.3 Local dislocation velocity profile at 200K (left) and 600K (right) along $\langle 1\bar{2}1 \rangle$ direction at different time. Segment positions with speed above 400 m/s are shown as red (on-line).

Fig.4 Contour plot the dynamical correlation of kinetic energy fluctuation at 200 K (left), at 600 K (right). The correlation is normalized by $4 \times 10^9/m^{*2}$, and the values are in the unit of $(m/s)^4$. Horizontal and vertical axes denote difference of dislocation position along $\langle 1\bar{2}1 \rangle$ direction and time difference, respectively.

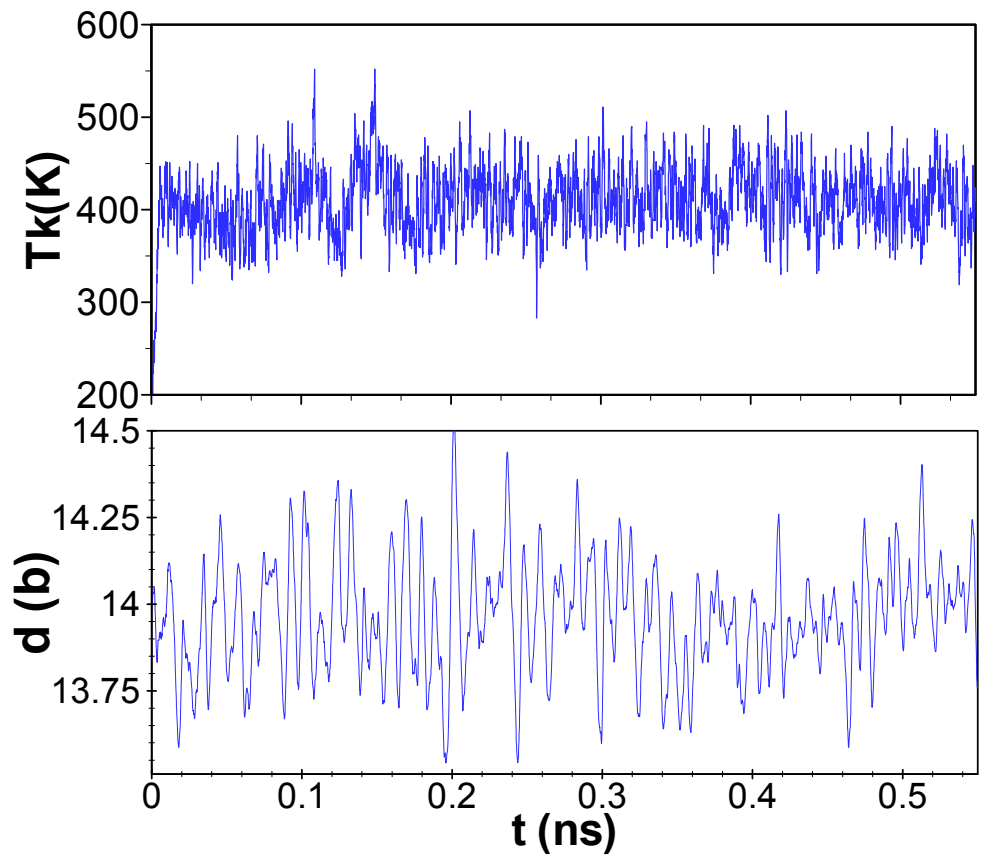

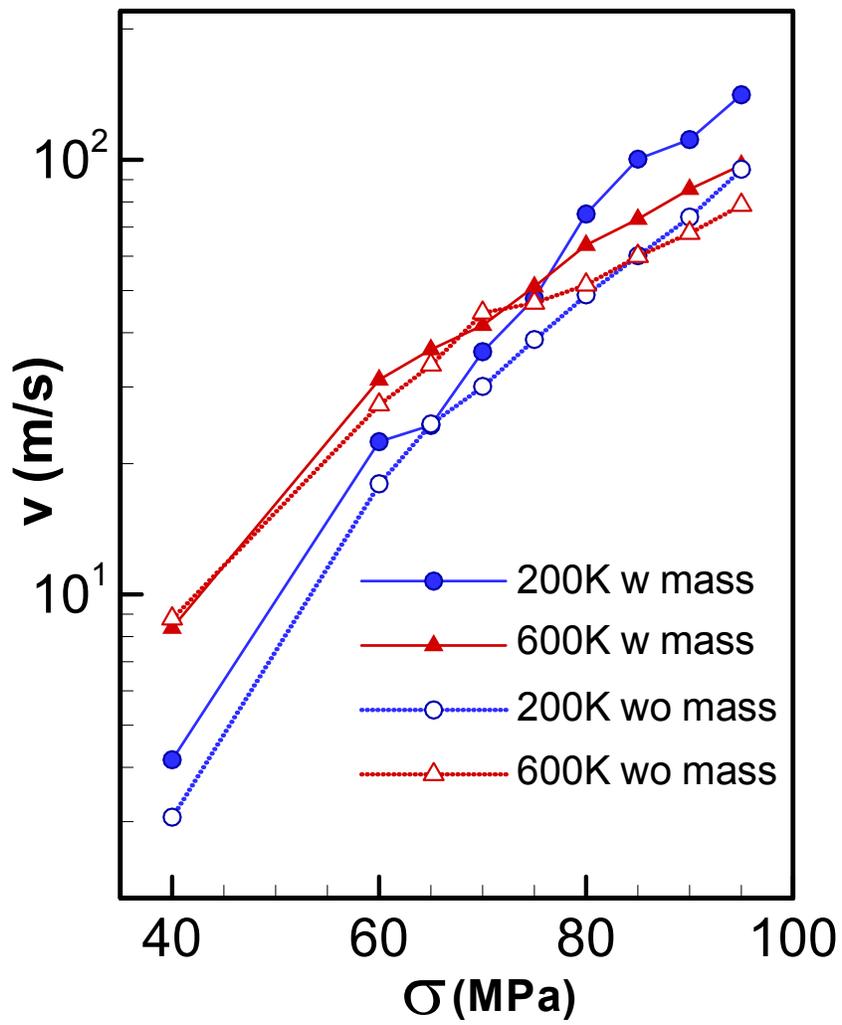

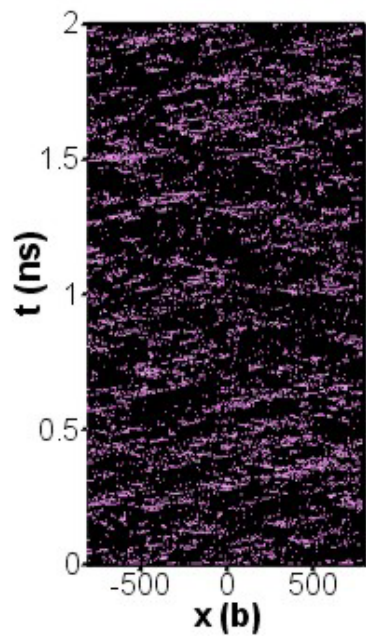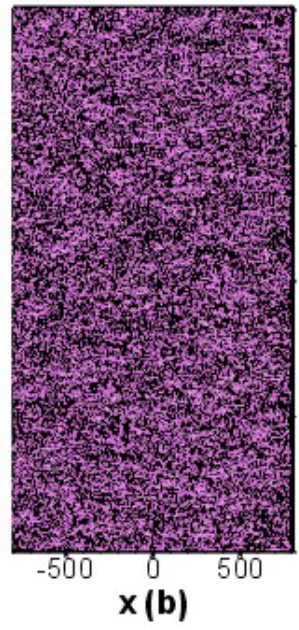

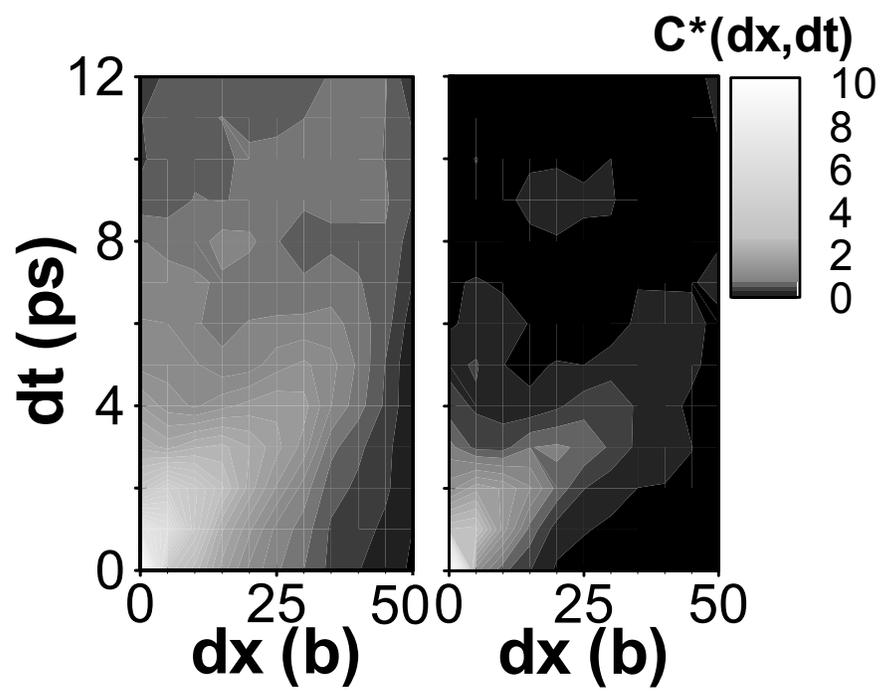